\documentclass[12pt]{article}
\usepackage[polish,english]{babel}
\usepackage[latin1]{inputenc}
\usepackage{times}
\usepackage[T1]{fontenc}
\usepackage{epsfig}

\begin{document}

\title{Compact Q-balls in the complex signum-Gordon model }

\author{H. Arod\'z $\;$ and $\;$ J. Lis\\$\;\;$ \\  Institute of Physics,
Jagiellonian University, \\ Reymonta 4, 30-059 Cracow, Poland}

\date{$\;$}

\maketitle

\begin{abstract}
We discuss Q-balls in the complex signum-Gordon model in  $d$-dimensional space for $d=1, 2, 3$. The Q-balls
have strictly finite size. Their total energy is a power-like function of the conserved $U(1)$ charge with the
exponent equal to $(d+2)(d+3)^{-1}$. In the cases $d=1$ and $d= 3$ explicit analytic solutions are presented.
\end{abstract}

\vspace*{2cm} \noindent PACS: 11.27.+d, 98.80.Cq, 11.10.Lm \\

\pagebreak

\section{ Introduction}

Q-balls are nontopological solitons which appear in certain nonlinear complex scalar field models \cite{1}, see
also \cite{2} for a review of early works and applications. Such solitons have been discussed in the context of
the dark matter problem, see, e.g., \cite{3}, and in condensed matter physics \cite{4}. Their static properties
as well as rather complex dynamics have been studied in several papers, see, e.g., \cite{5, 6}. Field potentials
in all considered models have a smooth behaviour close to their absolute minima. In consequence, the Q-balls
possess exponential tails and interact with each other.

Recently, a new class of scalar field models has been proposed, see \cite{7} for a review. Their common
characteristic feature is that the pertinent field potentials are not smooth at the absolute minima -- they are
V-shaped. It has turned out that models of this kind are quite interesting. First of all, they  have a sound
physical justification, and they are perfectly well-behaved from the physical viewpoint. Moreover, they can
support compact solitons (topological compactons), a whole variety of self-similar solutions \cite{8, 9}, as
well as time-dependent, finite energy excitations called oscillons \cite{10}. Despite rather unusual shape of
the field potentials exact, analytic solutions of all these types have been found. The complex signum-Gordon
model also has the V-shaped field potential: it is given by the formula $U(\psi) = \lambda |\psi|$, where
$\lambda >0$ and $\psi$ is a complex scalar field. The plot of $U(\psi)$   has an inverted conical shape. The
potential has its minimum right at the bottom of the cone.

In the present paper we show that the complex signum-Gordon model
possesses solutions with finite energy which are stationary
 -- the Q-balls. These Q-balls  are rather interesting for the following reasons. First, they appear rather
 surprisingly in the model with the very simple, conical field potential.
They are given explicitly as  exact solutions of the field equation. Other analytic Q-ball solutions are known
in literature, see, e.g., \cite{1, 6}, but our solutions seem to be simpler. The Q-balls have  a strictly finite
size. The scalar field, energy density and charge density vanish exactly outside certain ball of a finite
radius. In this sense, our Q-balls are compact (see also the second point  in the section Remarks). Another
distinguishing feature of our Q-balls is a very simple, power-like dependence of the total energy on the total
charge. We also show that there is a whole ladder of higher spherically symmetric Q-balls in the
three-dimensional space. Because of the compactness, one can also trivially combine several single Q-ball
solutions in order to obtain explicit solutions describing a collection of isolated Q-balls.

The plan of our paper is as follows. In Section 2 we present the field equation, and we discuss the energy and
charge of the Q-balls. Section 3 is devoted to explicit solutions of the field equation. In section 4 we have
collected several remarks.

\section{The field equation and general properties of the Q-balls }

Lagrangian of the complex signum-Gordon model has the  following form
\begin{equation}
L = \partial_{\mu}\psi^* \partial^{\mu} \psi - \lambda |\psi|,
\end{equation}
where $\psi$ is a complex scalar field in $(d+1)$-dimensional Minkowski space-time, $\lambda >0$ is a coupling
constant, $^*$ denotes the complex conjugation, $|\psi|$ is the modulus of $\psi$.   The field $\psi$, the
space-time coordinates $x^{\mu}$ and the constant $\lambda$ are dimensionless. Of course, in physical
applications they have to be multiplied by certain dimensional constants. Lagrangian (1) is invariant under the
global $U(1)$ transformations $\psi(x) \rightarrow \exp(i \alpha) \psi(x)$. The corresponding conserved charge
is given by the formula
\begin{equation}
Q = \frac{1}{2i} \int d^dx\: (\psi^*\partial_0\psi - \partial_0\psi^* \psi).
\end{equation}
The energy density
\begin{equation}
T_{00} = \partial_0\psi^* \partial_0\psi + \partial_i\psi^* \partial_i\psi + \lambda |\psi|
\end{equation}
has the absolute minimum at $\psi=0$.

The field potential $U(\psi) = \lambda |\psi|$ can be regarded as a limit of a regularized potential which is
smooth at $\psi=0$ and such that its second derivative $\partial^2 U/ \partial|\psi| \partial|\psi|$ taken at
$\psi=0$ diverges in that limit. As the example we take $ U_{\kappa}(\psi) = \lambda \sqrt{\kappa + |\psi|^2},$
where $\kappa$ is a positive constant. The signum-Gordon model is obtained in the limit $\kappa \rightarrow 0.$
The Euler-Lagrange equation in the regularized model has the form
\begin{equation}
\partial_{\mu}\partial^{\mu}\psi = - \frac{\lambda}{2} \frac{\psi}{\sqrt{\kappa + |\psi|^2}}.
\end{equation}
In the $\kappa \rightarrow 0$ limit the r.h.s. of this equation is equal to $- \lambda \psi/(2 |\psi|)$ if $\psi
\neq0$, and to 0 if $\psi =0$.

We use the standard Ansatz for symmetric Q-balls:
\[ \psi = F(r) \exp(i \omega x^0), \]
where $F$ is a real-valued function of  $r$, and $\omega > 0$ is a
real frequency. $r$ denotes the $d$-dimensional radial coordinate
in the cases  $d >1$ , and $r = x^1$ for $d=1$. It is convenient
to introduce the variable $y$ and the function $f(y)$:
\[ y = \omega r, \;\;\;\; f(y) = \frac{2\omega^2}{\lambda} F(r).
\] Equation (4) in the case of Q-balls in the limit $\kappa
\rightarrow 0$ acquires the form
\begin{equation}
f'' + \frac{d-1}{y} f' + f = \mbox{sign} f,
\end{equation}
where $'$ stands for $d/ dy$.  The $\mbox{sign}$ function has the
values $\pm 1$ when $ f \neq 0$ and $ \mbox{sign}(0)  = 0.$ Note
that Eq. (5) has the following constant solutions: $f = 0, \pm 1.$

Let us now briefly discuss essential mathematical aspects of Eq. (5). First, it is clear that $f$ has to satisfy
the condition $f'(0) =0$ if $d >1$. Furthermore, for physical reasons -- the finitness of the energy density (3)
-- $f$ has to be continuous and differentiable function of $y$. On the other hand, Eq. (5) implies that the
second derivative of $f$ has to be discontinuous at those points where $f$ changes its sign. Let $y_0$ be such a
point. Integrating both sides of Eq. (5) over a small interval around $y_0$ and shrinking that interval to the
point we find that $f'$ is continuous at $y_0$. The method to solve Eq. (5) consists of two steps. First, we
find solutions assuming that $f$ has a constant sign. In general, such solutions are valid on certain intervals
of the $y$ axis -- for this reason we call them partial.  In the second step we match such partial solutions
using the conditions of continuity of $f$ and $f'$. Differential equations with discontinuous terms are
well-known in mathematics and physics. Their general theory is based on the notion of so called weak solutions,
see, e.g., \cite{11, 12}. Our solutions fit in that framework.

We show in the next Section that there exist solutions of Eq. (5)
for which the charge $Q$ and the energy $E = \int d^dx\: T_{00}$
are finite. If $f(y)$ is such a solution, then formulas (2), (3)
give
\begin{equation}
Q = c_1 \frac{\lambda^2}{\omega^{d+3}}, \;\;\; E =  c_2 \frac{\lambda^2}{\omega^{d+2}}
\end{equation}
where $c_1, c_2$ are purely numerical constants which do not depend neither on $\omega$ nor $\lambda$. In the
$d=1$ case these constants are computed from the following formulas
\begin{equation}
c_1 = \frac{1}{4} \int^{\infty}_{-\infty} dy \: f^2(y), \;\;\;c_2 = \frac{1}{4} \int^{\infty}_{-\infty} dy \:
[f^2 + (\partial_y f)^2 + 2 |f|],
\end{equation}
while for $d >1$
\begin{equation}
c_1 = \frac{\Omega_d}{4}  \int^{\infty}_{0}\!\! dy \: y^{d-1} f^2(y), \;\;\;c_2 = \frac{\Omega_d}{4}
\int^{\infty}_{0}\!\! dy \: y^{d-1} [f^2 + (\partial_yf)^2 + 2 |f|],
\end{equation}
where $\Omega_1 = 2 \pi, \; \Omega_2 = 4 \pi$ ($\Omega_d = \int d
\Omega_d$, where $d \Omega_d$ is the $d$-dimensional solid angle
element). We shall see in the next Section that the frequency
$\omega$ and the radius $r_0$ of the Q-balls are related, $  r_0=
y_0/\omega$, where $y_0$ is a numerical constant. Therefore,
formulas (6) imply that $Q \sim r_0^{d+3}, \; E \sim r_0^{d+2}$.

Formulas (6) imply that for a fixed solution $f$ of the rescaled
radial equation (5) the total energy and the charge are related by
the very simple formula
\begin{equation}
E = c_2 \: \lambda^{\frac{2}{d +3}} \: \left(\frac{Q}{c_1} \right)^{\frac{d+2}{d+3}}.
\end{equation}
The power-like dependence of $E$ on $Q$ is the consequence of
power-like dependence on $\omega$  in formulas (6). The latter one
can be related to a scale invariance of the field equation (4) in
the limit $\kappa \rightarrow 0$, \cite{8, 9}.

Because the exponent on the r.h.s. of formula (9) is smaller that 1, one spherical Q-ball with the total charge
$Q= Q_1 + Q_2$ has smaller energy that two smaller spherical Q-balls with the charges $Q_1>0$, $Q_2>0$. In other
words, it is energetically favourable for the two Q-balls to merge.

\section{The explicit form of the solutions}

\subsection{The d=1 case}
In this case Eq. (5) has a family of partial solutions of the form
\[
f_+(y) = 1 + \alpha \sin y,
\]
where $\alpha$ is a real constant. They obey Eq. (5) on (sub)intervals of the $y$-axis determined by the
condition $ f_+ > 0$. Using the translations  $y \rightarrow y - a$  one can trivially generate further
solutions. Below we will discuss in detail only  the simplest solutions, omitting those which can be obtained by
the space or time translations, or by Lorentz boosts.

The basic one dimensional Q-ball solution $f_0(y)$ is obtained by combining the trivial solution $f=0$ with the
solution $f_+$. The matching conditions discussed in the previous Section imply that $\alpha =-1$, and that the
two solutions match each other at the points $y= \pi/2, \; 5\pi/2$. Hence,
\begin{equation}
f_0(y)=\left\{
\begin{array}{lcl} 0 & \;\;\; \mbox{if} \;\;\; & y \leq \pi/2, \\ 1 - \sin y & \;\;\; \mbox{if} \;\;\;
&  \pi/2 < y < 5\pi/2, \\   0 & \;\;\; \mbox{if} \;\;\; & y \geq 5 \pi/2.
\end{array} \right.
\end{equation}
Formulas (7) give $c_1= 3\pi/4, \; c_2= 2 \pi.$

The basic solution (10) gives rise to a whole family of multi-Q-ball solutions. They are obtained just by
summing translated or Lorentz boosted solutions $ f_0(y)$. The only restriction is that the supports of the
$1-\sin(y-a)$ parts should not overlap. Such separate Q-balls do not interact with each other.

The $d=1$ complex signum-Gordon model was considered in \cite{8}
in connection with a model of a string in a three dimensional
space, pinned to a straight line coinciding with the $x^1$ axis.
In that model $Re \psi, \; Im \psi$ are just the two coordinates
giving the deviation of the string from the pinning straight line.
The Q-ball solutions presented above describe the string rigidly
rotating around the $x^1$ axis with constant angular velocity
equal to $\omega$.

\subsection{The d=3 case}

Let us start from a particular solution of the form
\begin{equation}
f_+(y) = 1 + \alpha \frac{\sin y}{y} + \beta \frac{\cos y}{y},
\end{equation}
where $\alpha, \beta $ are real constants. It obeys Eq. (5) in the
(sub)intervals determined from the condition $f_+(y) >0$. The
simplest Q-ball solution is composed of the trivial solution $f=0$
in the region $y \geq y_0$ and $ f_+$ in the region $ 0 \leq y
\leq y_0$. The condition $ f'_+(0)=0$ is satisfied only when
$\beta =0$. The matching conditions at $y=y_0$ have the form
$f_+(y_0)=0$, $ f'_+(y_0) =0$. They give $ \alpha = - y_0/\sin
y_0$ and the following equation for $y_0$
\[
\tan y_0 =y_0. \] This equation has infinitely many solutions, but
only for the first one, i.e.,
\[ y_0 \approx 4.4934, \]
the function $f_+$ has positive values in the whole interval $[0,
y_0)$. Thus, the basic 3-dimensional Q-ball solution is given by
\begin{equation}
f_b(y)=\left\{
\begin{array}{lcl}  1 - \frac{y_0}{y} \frac{\sin y}{\sin y_0} & \;\;\; \mbox{if} \;\;\;
&  0 \leq y < y_0, \\   0 & \;\;\; \mbox{if} \;\;\; & y \geq y_0.
\end{array} \right.
\end{equation}
The constants $c_1, c_2$ have the following values $ c_1 = 5 \pi
y_0^3/6, \; c_2 = 2 \pi y_0^3$.

Equation (5) can formally be regarded as Newton's equation of motion for a fictitious particle in the effective
potential $V_{eff} = f^2/2 - |f|$ and with the friction force equal to $-(d-1)f'/y$.  The role of time is played
by $y$, while $f$ gives the position of the particle. Such reinterpretation of radial equation was used already
by, e. g., Lee and Coleman  \cite{1}. In the case of Q-ball solution, the particle starts at the time $y=0$ from
a certain point $f
>0$ with vanishing velocity $f'(0)=0$, reaches the point $f=0$ at the time $y_0$, again with vanishing velocity
$f'(y_0)=0$, and rests at that point for later times ($y > y_0$). It is clear that also other trajectories are
possible. In particular, the particle can have an excess of energy when arriving to the point $f=0$, hence it
continues to move with the negative velocity $f'(y)$ to the negative values of $f$, bounces back and finally
arrives at the point $f=0$ with vanishing velocity where it stops. This is possible because the particle looses
its energy due to the friction. This trajectory corresponds to a higher Q-ball solution which is composed of the
following three partial solutions
\[ f_+(y) = 1 - \alpha_0 \frac{\sin y}{y}, \;\;   f_-(y) = -1 + \alpha_1 \frac{\sin y}{y}
+ \beta_1 \frac{\cos y}{y}, \;\; f=0. \] The  matching conditions
$ f_+ = f_-, \: f'_+ = f'_-$ at $y_{10}$, and $f_-=0, \: f'_-=0 $
at $y_{11} > y_{10}$ give
\[ \alpha_0 = \frac{y_{10}}{\sin y_{10}}, \;\; \alpha_1 = y_{11} \sin y_{11} + \cos y_{11},
\;\; \beta_1 = y_{11} \cos y_{11} - \sin y_{11},  \]
\[
y_{10} \approx 3.4826, \; y_{11} \approx 8.4970. \] This solution has single isolated zero  at $y=y_{10}$, and
it merges with the $f=0$ solution at $y_{11}$.  The $f_+$ part is for $ y \in [0, y_{10}]$ and the $f_-$ part
for $y \in [y_{10}, y_{11}]$.

It is clear that the fictitious particle can also bounce back $n
>1$ times -- such a trajectory corresponds to the Q-ball solution with $n$
isolated zeros located at $y_{n0}, ... y_{nn}$.

Similarly as in the $d=1$ case, single Q-balls can be translated
and boosted. One can also simply add solutions with non
overlapping supports of the $f_+, f_-$ parts in order to form
noninteracting multi-Q-ball solutions.

\section{Remarks}

\noindent 1. We have omitted a discussion of the case $d=2$
because  Q-ball solutions in two spatial dimensions are completely
analogous to the ones from the $d=3$ case. The only difference is
that  the functions $\sin y /y$, $ \cos y/y$ are now replaced by
the Bessel functions $J_0(y), \: Y_0(y)$, respectively, with very
similar plots \cite{13}. Therefore, the structure of solutions is
essentially the same as in the $d=3$ case. This can also be seen
from the mechanical interpretation of the radial equation (5).

\noindent 2. Our Q-ball solutions approach the vacuum field $\psi=0$ exactly at the radius $r_0 = y_0/\omega$.
For $r \rightarrow r_0-$ the profile function $F$ has a parabolic shape, as can  be seen from formulas (10) and
(12). Such behaviour is typical for field models with V-shaped interaction terms \cite{7}. The energy of the
Q-balls is strictly localized inside the ball of radius $r_0$ -- there is no exponential tail.

It should be noted that a compact Q-ball was found already by Werle \cite{1}. In that paper a rather peculiar
field potential which involves two fractional powers of $|\psi|$ was considered. Moreover, the first derivative
of the potential becomes infinite for the vacuum field.

\noindent 3.  In general, Q-ball dynamics is very complex already on the level of purely classical theory, see,
e.g., \cite{5}, and still richer when one includes quantum effects, see, e.g., \cite{14} and references therein.
The simplicity of the Q-balls of the complex signum-Gordon model can perhaps facilitate the theoretical studies,
and make them interesting for applications. Nevertheless, we do not expect that their dynamics will be much
simpler. To illustrate the point, let us take the problem of stability of the Q-balls.  We have checked that the
basic Q-balls are stable against just radial shrinking or expanding. However, a slightly perturbed non symmetric
Q-ball can apriori behave in various ways. For example, it can  perhaps decay by emitting a number of small
Q-balls, or by emitting waves of the scalar field. Undoubtedly, there are many interesting problems in the
dynamics of Q-balls to be investigated.

\section{Acknowledgement}
This work is supported in part by the project SPB nr. 189/6.PRUE/2007/7.

\end{document}